\documentclass{mn2e}
\usepackage{amsmath, amssymb, epsfig}
\def\simlt{\lower.5ex\hbox{$\; \buildrel < \over \sim \;$}}
\def\simgt{\lower.5ex\hbox{$\; \buildrel > \over \sim \;$}}
\def\simpt{\lower.5ex\hbox{$\; \buildrel \propto \over \sim \;$}}

\def\kms{\mbox{ km s$^{-1}$}}

\def\ciii{C\,{\scriptsize III}]~}
\def\aliii{Al\,{\scriptsize III}~}
\def\civ{C\,{\scriptsize IV}~}
\def\mgii{Mg\,{\scriptsize II}~}

\def\feii{Fe\,{\scriptsize II}~}
\def\alii{Al\,{\scriptsize II}~}
\def\siv{Si\,{\scriptsize V}~}
\def\siiii{Si\,{\scriptsize III}]~}
\def\heii{He\,{\scriptsize II}~}
\def\hb{H$\beta$ }
\def\lya{Ly$\alpha$ }

\begin{document}

\title[The role of black hole mass in quasar radio activity]
{The role of black hole mass in quasar radio activity}
\author[R.B. Metcalf \& M. Magliocchetti]
{R.B. Metcalf$^{1}$ \& M. Magliocchetti$^{2}$ \\ 
$^1$ Hubble Fellow, Department of Astronomy and Astrophysics, University of 
California, High Street, Santa Cruz, CA95064\\
$^{2}$ S.I.S.S.A., Via Beirut 2-4, 34014 Trieste, Italy } 

\maketitle

\vspace {7cm}

\begin{abstract}
We use a homogeneous sample of $\sim 300$, 
$0.3\simlt {\rm z}\simlt 3$, radio-loud quasars drawn from the FIRST and 2dF 
QSO surveys to investigate a possible 
dependence of radio activity on black-hole mass. By analyzing composite 
spectra for the populations of radio-quiet and radio-loud QSOs -- chosen to 
have the same redshift and luminosity distribution -- we find 
with high statistical significance 
that radio-loud quasars are on average associated with black holes of masses 
$\sim 10^{8.6}{\rm M_\odot}$, about twice as large as those measured for 
radio-quiet quasars ($\sim 10^{8.3}{\rm M_\odot}$).

We also find a clear dependence of black hole mass on optical luminosity of 
the form ${\rm log \left(\frac{M_{BH}}{M_\odot}\right)_{RL}}= 8.57(\pm0.06) -
0.27(\pm 0.06) ({\rm M_B} + 24.5)$ and
${\rm log \left(\frac{M_{BH}}{M_\odot}\right)_{RQ}}= 8.43(\pm0.05) -
0.32(\pm 0.06) ({\rm M_B} + 24.5)$, respectively for the case of radio-loud 
and radio-quiet quasars. It is intriguing to note that these two trends run 
roughly parallel to each other, implying that radio-loud quasars are associated to 
black holes more massive than those producing the radio-quiet case 
{\it at all sampled luminosities}.
On the other hand, in the case of radio-loud quasars, we find evidence 
for only a weak (if any) dependence of the black hole mass on radio power.
The above findings seem to support the belief that there exists -- at
a given optical luminosity -- a threshold black hole mass associated
with the onset of significant radio activity such as that of
radio-loud QSOs; however, once the activity is triggered, there
appears to be very little connection between black hole mass and level
of radio output.
\end{abstract}


\section{Introduction}\label{sec:introduction}

Despite the recent progress made in understanding the cosmological 
evolution of quasars and their host galaxies (e.g. Silk \& Rees 1998; 
Cavaliere \& Vittorini 2002; Dunlop et al. 2003; 
Granato et al. 2004; McLure \& Dunlop 2004; Di Matteo, Springel, Hernquist 
2005) there is still little known about the processes that produce
the sub-class of radio-loud (RL) AGN which exhibit radio activity from powerful relativistic jets. 

Neither host galaxy morphology nor large-scale environment seem to be a
major determinant of a quasar's radio activity
 as the two populations (or at least their brighter counterparts) 
are found in the same kind of bulge-dominated spheroidal galaxies 
(e.g. McLure et al. 1999; Schade, Boyle \& Letawsky 2000; Dunlop et al. 
2003) and tend to inhabit groups and clusters of similar richness (e.g. 
McLure \& Dunlop 2001; Wold et al. 2001).   
At face value, these findings imply that enhanced jet activity 
from AGNs is associated with the pc/sub-pc scale behavior of a galaxy, 
and more specifically related to some properties of its central black 
hole.
The extraction and collimation of energy from the accretion
disk to form jets most likely involves some magnetohydrodynamic
mechanism (Blandford \& Payne 1982) and/or the spin of the black-hole
(Blanford \& Znajeck 1977; Blandford 2002). In addition, the mass of
the central black hole could play an important role in regulating the 
radio-loud/radio-quiet dichotomy. 

The issue of a possible dependence of radio activity on black hole mass 
has recently been the subject of many scientific debates.
A number of authors (e.g. Laor et al. 2000; McLure \& Dunlop 2002; Dunlop et 
al. 2003; Marziani et al. 2003; McLure \& Jarvis 2004) have concluded that 
radio-loud sources 
(both quasars and radio-galaxies) are systematically associated with black 
holes of greater mass than their radio-quiet (RQ) counterparts. The proposal, 
first made by Laor (2000), that there exists a threshold black hole
mass below which a quasar cannot be radio-loud 
was further confirmed by the clustering results of 
Magliocchetti et al. (2004) who find local (${\rm z\simlt 0.3}$) 
radio galaxies to be powered by 
black holes with masses ${\rm M_{BH}\simgt 10^{9} M_\odot}$, about an order of 
magnitude higher than those found for radio-quiet quasars by means of the 
same statistical analysis (see e.g. Porciani, Magliocchetti \& Norberg 
2004). Furthermore, some author also claim that there is a continuous 
relationship 
(although with a large scatter) between radio-power and black hole
mass that holds over about a decade in radio power, from 
local 'inactive' galaxies (Franceschini, Vercellone \& Fabian 1998), to 
powerful radio-galaxies (McLure et al. 2004) and quasars (e.g. Lacy et al. 
2001; McLure \& Jarvis 2004).

On the other side, other authors have argued strongly against any  
dependence of radio luminosity on black hole mass and the 
existence of a threshold mass above which the fraction of 
radio-loud vs radio-quiet sources drastically increases; in fact, they find 
evidence for radio-loud objects associated with black holes with masses as
low as  ${\rm 10^6 M_\odot}$ (e.g. Oshlack et al. 2002;  Ho 2002; Woo
\& Urry 2002a,b; Snellen et al. 2003. See however the dissenting view of 
Jarvis \& McLure 2002 who re-analysed the Oshlack et al. 2002 sample by taking into 
account corrections due to Doppler boosting effects and the geometry of the 
broad-line region).  

The aim of this work is to reexamine the issue of radio 
activity vs. black hole 
mass by investigating the spectral properties of a new, homogeneous sample 
of $\sim 300$ radio-loud quasars -- spanning the redshift range 
$0.3\simlt {\rm z}\simlt 3$ -- obtained by the joint use of the FIRST and 
2dF Quasar Redshift Surveys. 
Average virial mass estimates are derived for these objects and 
compared to those of a similar-size sample of radio-quiet quasars,
also drawn from the 2dF QSO survey.  The two samples are selected to
have the same redshift and 
magnitude distributions to avoid any biases due to cosmological evolution 
and/or selection. We also present estimates of 
black hole mass as a function of optical and radio luminosities for
these sources.
Throughout this paper we will assume $\Omega_0=0.3$, $\Lambda=0.7$, 
$\rm h_0=0.7$ as the latest results
from the joint analysis of CMB and 2dF data indicate (see e.g. Lahav
et al. 2002).

\section{Data}\label{sec:data}
For our analysis we make use of the catalogue of 
radio-loud quasars presented in Cirasuolo et al. (2003a) and 
Cirasuolo, Magliocchetti \& Celotti (2005). Briefly, a catalogue of 352 
sources has been obtained by matching together objects
from the FIRST (Faint Images of the Radio Sky at Twenty centimeters) survey 
(Becker et al. 1995) and the 2dF QSO redshift survey (Croom et al. 2004). 

The FIRST survey includes 811,117 sources (April 2003 release used for this 
work) observed at 1.4~GHz down to a flux limit S$_{\rm
  1.4GHz}\simeq$0.8~mJy and is substantially complete. Its 
completeness has been estimated to be 95 per cent at 2~mJy and 80 
per cent at 1~mJy. The survey covers a total of about 9033 square degrees on 
the sky (8422 square degrees in the North Galactic cap and 611 in the South 
Galactic cap). Point sources at the detection limit of the catalogue have 
positions accurate to better than 1 arcsec (90 per cent confidence level).

The 2dF QSO redshift survey (2QZ) includes $\sim$21,000, $18.25\le 
{\rm b_J}\le 20.85$, quasars with reliable spectra and redshift 
determinations covering two $75^\circ \times 5^\circ$ declination strips 
centered on $\delta=-30^\circ$ (South Galactic cap) and $\delta=0^\circ$ 
(North Galactic cap). In order to guarantee a large photometric completeness 
($> 90$~per cent) for quasars within the redshift $0.3\le {\rm z}\le 2.2$, 
the following color selection criteria were applied: $(u-b_j)\le 0.36$; 
$(u-b_j)< 0.12-0.8\;(b_j-r)$; $(b_j-r)<0.05$.
Spectroscopic observations of the input catalogue were made with the
2-degree Field (2dF) instrument at the Anglo-Australian Telescope. The
spectra were classified both via cross-correlation with specific
templates (AUTOZ, Croom et al. 2004) and by visual inspection.

The FIRST and 2QZ surveys overlap in the region $ 9^h \; 50^m
\leq {\rm RA(2000)} \leq 14^h \; 50^m$ and $ -2.8^{\circ} \leq {\rm
Dec(2000)} \leq 2.2^{\circ}$.  In this region there are 10,110
optical quasars from the 2QZ and $\sim$45,500 radio sources down to
S$_{\rm 1.4 GHz}$=1~mJy from the FIRST catalogues.  Not all these radio
sources correspond to separate objects.  As described in Cirasuolo et
al. (2003a), the authors used the algorithm developed by Magliocchetti
et al. (1998) to collapse multi-component sources into single objects
having radio fluxes equal to the sum of the components.  
All the optical-radio pairs with offsets of less than 2 arcsec are
considered true optical identifications.  The value of 2 arcsec was
chosen, after a careful analysis, as the best
compromise between maximizing the number of real associations (estimated to
be $\sim 97$ per cent) and minimizing the contribution from spurious
identifications (a negligible 5\%) (Magliocchetti \&
Maddox 2002). In addition, in order to verify the reliability of the
associations, all the radio-optical pairs obtained from the collapsing
algorithm were checked by eye on the FIRST image cutouts.

This procedure has lead to a homogeneous sample 352 quasars (113 presented in 
Cirasuolo et al. 2003a, while the remaining 239 given in Cirasuolo et al. 
2005) with good redshift
determinations from the 2QZ and with radio fluxes ${\rm
S_{1.4GHz} \ge 1}$ mJy over an effective area of 284 square
degrees. 
Due to the 2QZ and FIRST flux limits 
all these sources can be considered radio-loud whether one uses the 
criterion of a radio-to-optical ratio $R^{\*}_{\rm 8.5GHz}\simgt 10$ 
(corresponding to  $R^{\*}_{\rm 1.4GHz}\simgt 30-40$ for objects with a 
spectral index $\alpha_{\rm R}=0.8$; Kellermann et al. 1989) or 
a threshold of $\rm log_{10}P_{5GHz}\sim 24$~[W~Hz$^{-1}$sr$^{-1}$]
as used by Miller et al. (1990).   

The above radio-quasar sample corresponds to about 3.5~per cent of the original 
2dF quasar population. This value is somewhat lower than what found in previous works 
($\sim$ 10-20~per cent; White et al. 2000; Hewett, Folz \& Chaffee 2001; Ivezic et al. 
2002). As demonstrated by Cirasuolo et al. (2003a; 2003b), the reason for this finding originates from 
the fact that the radio-loud fraction is a function of absolute optical luminosity so that the low 
radio-loud fraction of the 2QZ/FIRST sample is due to the relative optical faintness of the 2QZ quasars.

Luminosities for the above sample were obtained by converting magnitudes 
from the the $b_{\rm J}$ to the B band.
The mean $B-b_{\rm J}$ was computed from the composite quasar spectrum
compiled by Brotherton et al. (2001) from $\sim 600$ radio-selected
quasars in the FIRST Bright Quasar Survey (FBQS). The K-correction in the B 
band has also been computed from the
Brotherton et al. (2001) composite quasar spectrum.

\begin{figure}
\centering\epsfig{figure=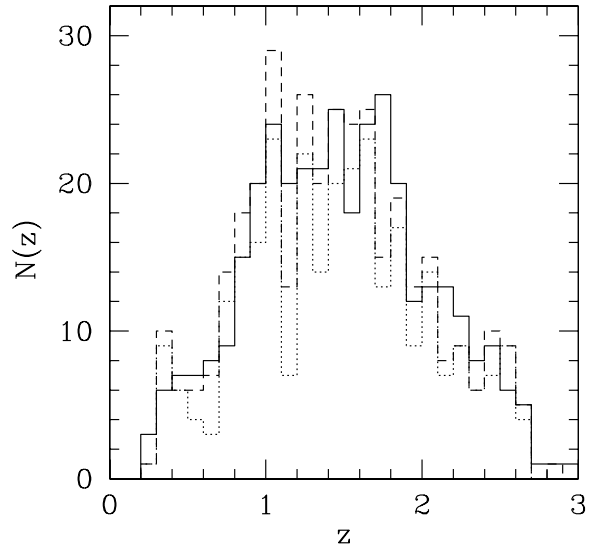,height=3.5in}
\caption {Redshift distribution for the samples under examination. 
The solid line is 
for radio-quiet sources, the dashed line for radio-loud sources, and the 
dotted line represents those radio-loud quasars with available optical 
spectrum from the 2dF QSO survey.}\label{nz_comp_new}
\end{figure}

\begin{figure}
\centering\epsfig{figure=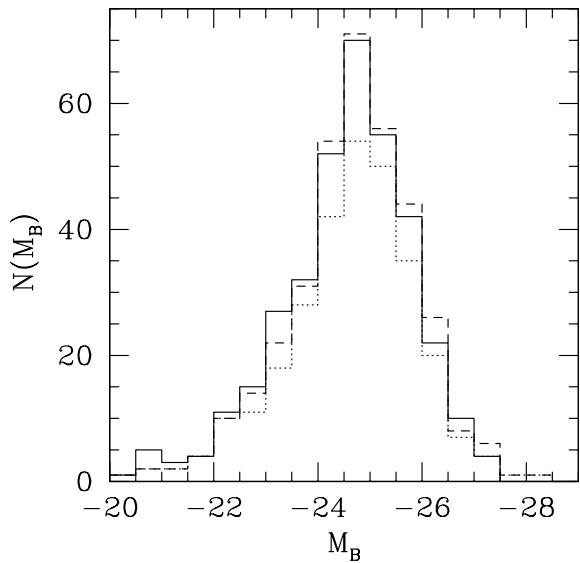,height=3.5in}
\caption[]{Luminosity distribution for the samples under examination. 
The line coding is as in figure \ref{nz_comp_new}.}\label{nmag_comp_new}
\end{figure}

Since the aim of this work is to analyze the (optical) spectral differences 
between radio-loud (RL) and radio-quiet (RQ) quasars, we have
generated a sample of 352 radio-quiet QSOs by extracting sources at random 
from the 2QZ survey in the same area covered by our radio-loud catalogue.
These sources were chosen to have the same redshift and magnitude 
distributions as the radio-loud sample (see figures \ref{nz_comp_new}, 
\ref{nmag_comp_new}) in order to exclude any possible bias due to 
cosmological and luminosity evolution and/or selection in our analysis. 
    The application of the one-dimensional Kolmogorov-Smirnov (KS) test finds 
     that both the optical luminosity and the redshift distributions
     of the RL and RQ samples are indistinguishable: the probabilities
     of getting  larger KS statistics are p=0.99 and p=0.76 respectively. These
     probabilities are so high because the RQ QSOs have been picked by
     hand to have the same redshift distribution as the RL QSOs.  This
     also confirmed in the case of the z-$\rm M_{B}$  distribution;
     application of the two-dimensional KS test gives p=0.82.

Optical spectra for both radio-loud and radio-quiet quasars were 
subsequently obtained from the 2dF QSO survey public release 
(http://www.2dfquasar.org).  Sixty one of the radio-loud 
sources in our catalogue did not have spectra available in the
archive, leaving 291 that did.  
 It is likely that the reason for this incompleteness arises from 
     not having included (as their spectra are not publicly available) from the 2QZ survey those 
     sources belonging to
     the Brotherton et al. (1998) sample. We have therefore made sure
     that when the radio-loud sources without available spectra are
     included in the sample the redshift and luminosity distributions are still
     consistent with those of the radio-quite sample.  The
     one-dimensional KS test gives p=0.82 and p=0.52 for the redshift
     and magnitude distributions respectively and the two-dimensional
     KS test gives p=0.71 -- all very consistent with coming from the same
     distribution. 
Figures \ref{nz_comp_new} and \ref{nmag_comp_new} show these distributions, 
where the dashed histograms indicate the total radio-loud sample, while
the dotted ones show the portion of radio-loud
sample with available optical spectra. Based on the above results we
are confident that the two sets represent the same underlying
population and in the following analysis we will use the sub-sample of
291 radio-loud quasars to compare their spectral properties  
with those of radio-quiet sources.

\section{Constructing Composite Spectra}\label{sec:constr-comp-spectra}

\begin{figure*}
\centering\epsfig{figure=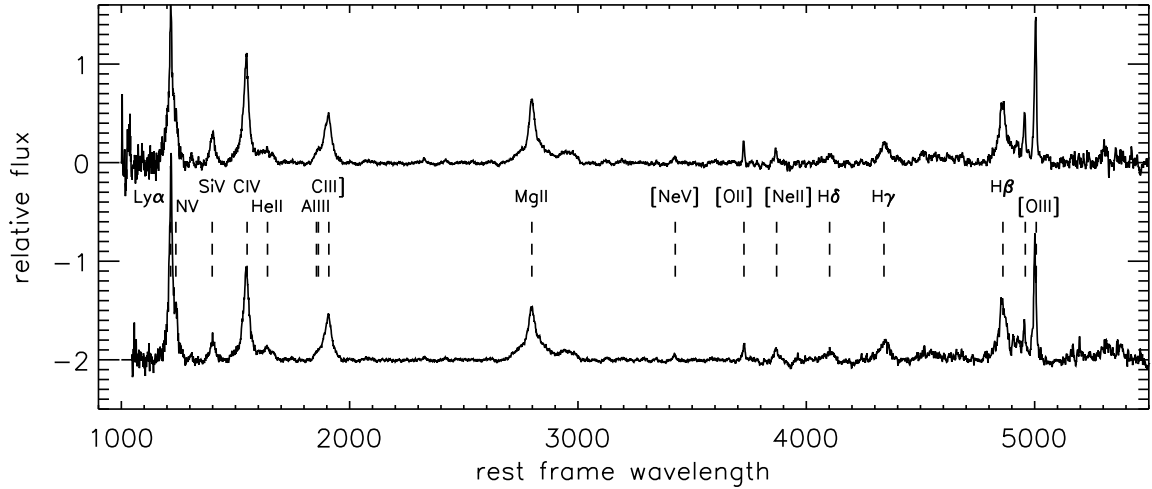,height=4.75in}
\vspace{-2 cm}
\caption[]{Continuum normalized composite spectra for our
  sample of radio-quite (on top) and radio-loud (on bottom) QSOs.}
\label{fig:composite}
\end{figure*}

\begin{figure*}
\centering\epsfig{figure=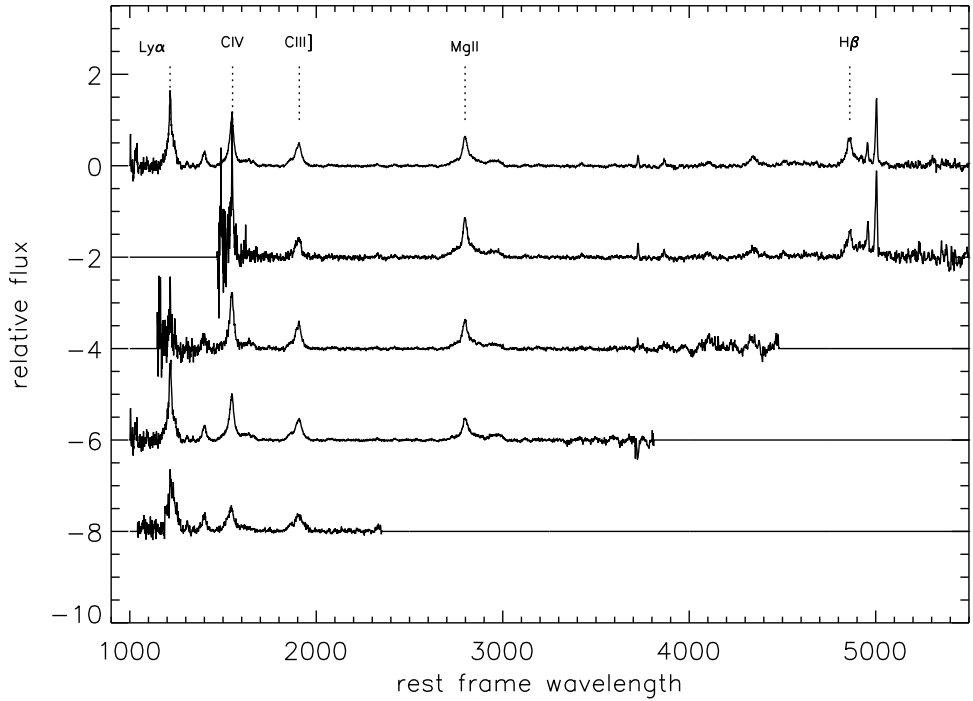,height=4.0in}
\caption[]{Composite spectra for the sample of radio-quiet QSOs
binned in B-band absolute magnitude.  The M$_{\rm B}$ ranges for the spectra
from bottom to top are: -27 to -30, -27 to -25, -25 to -24 and -21 to
-24~mag.  The very top curve  is for all the RQ QSOs and is 
the same spectrum as in figure~\ref{fig:composite}.  From the
spectral ranges it is possible to see both the redshift dependent selection
effects and the cosmological luminosity evolution of QSOs.}
\label{composite_RQ}
\end{figure*}

\begin{figure*}
\centering\epsfig{figure=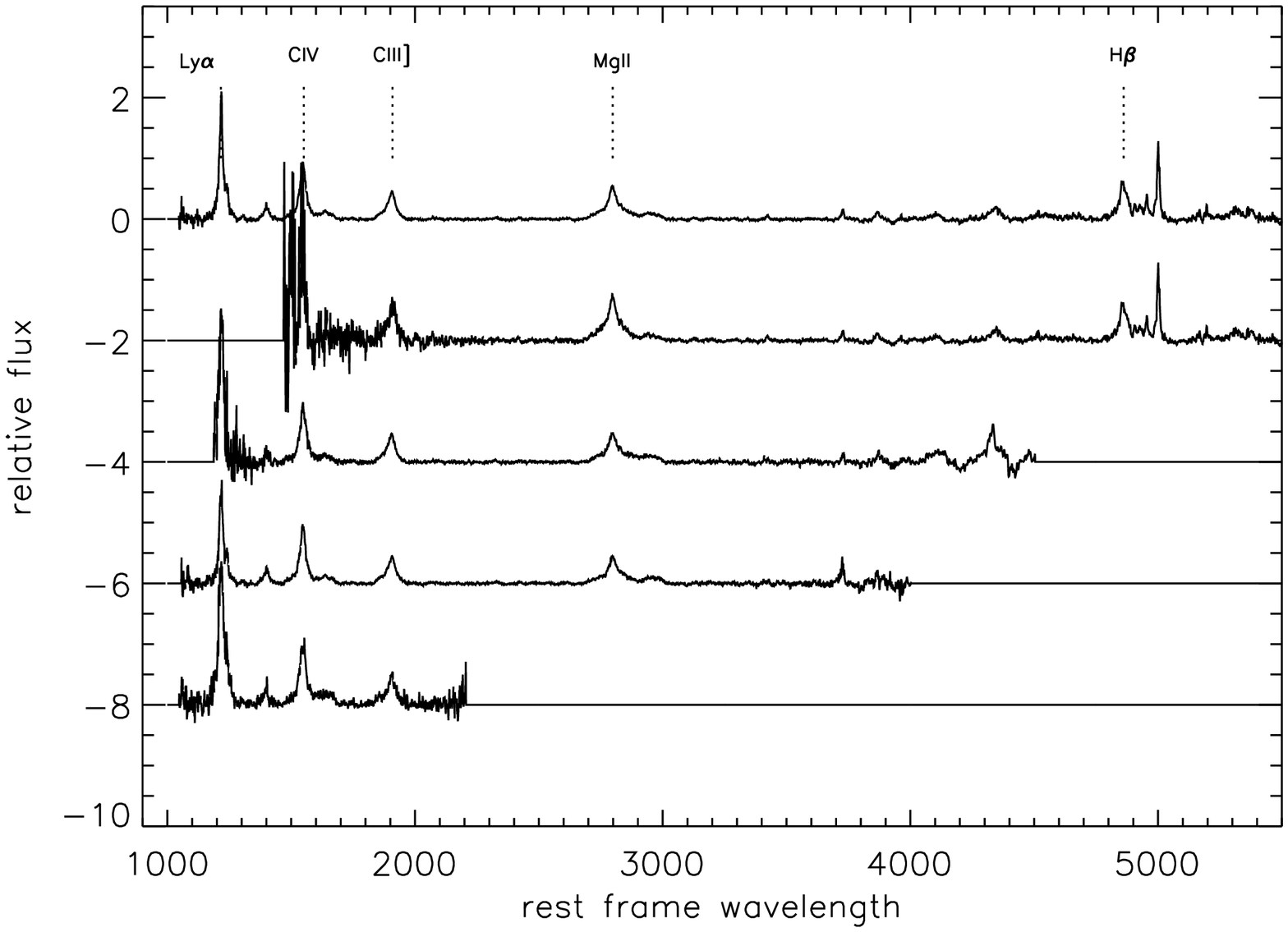,height=4.0in}
\caption[]{Same as figure~\ref{composite_RQ} only for
  radio-loud QSOs.}
\label{composite_RL}
\end{figure*}

The typical 2dF spectrum has a dispersion of 4.3~\AA~per pixel and a 
resolution of $\simeq 9$~\AA~over the range 3700-7900~\AA. The median 
signal-to-noise ratio is $\sim$5~pixel$^{-1}$ (Croom et al. 2002). 
This noise level is too large 
to accurately measure individual line widths.  Instead we must rely on
composite spectra to reveal spectral  
properties of the population as a whole with higher signal to noise.

The composite spectra are constructed by performing the following steps on each
of the contributing spectra.  First, cosmic rays and bad pixels are
removed. Then, since 2dF spectra are not flux-calibrated, we normalized 
each spectrum to a local continuum level. The continuum spectrum is calculated 
by removing the regions in the spectrum that contain known emission
features and then interpolating the spectrum over these regions.  
The interpolation was done by taking the median of the spectrum in two
portions of spectra on either side of each line region and then linearly 
interpolating
between them. The wavelength ranges used for the spectral features
and the adjacent interpolation regions are the same as in Croom et al. (2002). 
We did a further 50 pixel wide median smoothing which we judge to be the best 
compromise between minimize the noise in our 
continuum spectrum and at the same time preserve its large-scale features.  
The original spectrum was then divided by the continuum spectrum.
This continuum-normalized spectrum is then shifted
to the rest frame-wavelength according to the redshift measured by the
2dF survey and then it is linearly interpolated at intervals of
every 1~\AA.  The final composite spectrum is made by taking 
the median of these individual QSO spectra at each 1~\AA~interval or ``pixel''.

The composite spectra for all the radio-loud and radio-quiet QSOs are
shown in figure~\ref{fig:composite}. 
As a result of the spectral range of the 2dF spectrograph, no single QSO 
contributes to the full wavelength range of the spectra shown here.
It can be seen that the noise level 
varies over the length of the spectra, becoming largest at either
end, because the number of contributing spectra changing with 
wavelength.

The QSO samples are further divided into four $\rm M_B$ bins and the
composite spectra are calculated and shown in
figures~\ref{composite_RQ} and \ref{composite_RL}. The numbers of quasars 
contributing to each composite spectrum at the different magnitudes are: 
73 RL and 92 RQ quasars in the $-21\le \rm M_B < -24$ range; 96 RL and 122 RQ in the 
$-24\le \rm M_B < -25$ range; 112 RL and 129 RQ in the $-25\le \rm M_B < -27$ range 
and 6 RL and 4 RQ in the $-27\le \rm M_B < -30$ range. Note that the
luminosity interval for which a single line can be measured is 
limited.  Any further division of the samples in $\rm M_B$ tends to
increase the noise to such a level that comparisons of the RL and RQ
composite spectra are not very meaningful.

One might be concerned that coarsely binning the sample in luminosity
might obscure the meaning of any differences between the RL and
RQ composite spectra -- the differences could be caused by QSOs of the
same luminosity or different luminosities.  However, it can easily be
shown that if the RL and RQ samples have the same luminosity distribution
then any difference in their composite spectra must be caused by
differences in QSOs of the same luminosity (see appendix).  As 
shown in section~\ref{sec:data}, our samples are consistent with
having the same luminosity distribution.  In addition, the bootstrap
method for estimating errors on the line widths, discussed in the next
section, should take into account any coincidental differences in the
luminosity distributions.  For these reasons it makes sense to use
composite spectra in coarse luminosity bins to compare the black hole
masses of the RL and RQ quasars.

In addition, it has been shown that more generally radio
emission does not exhibit any tight correlation with visible luminosity 
(e.g. Cirasuolo et al. 2003b), since sources with a particular optical 
luminosity can have radio powers spanning up to three orders of 
magnitude.  This implies that our sample is representative in this
respect.

It is probable that there are other ``hidden'' properties that are
correlated with both radio emission and particular spectral
properties, either because of a common cause or a direct causal connection.
There have been claims that various spectral 
properties of QSOs depend on their position on the so-called Eigenvector 1 
(E1) 
parameter space, which relates the full width at half maximum (FWHM) of the 
broad component of H${\beta}$ (H${\beta}_{\rm BC}$) with the ratio between 
the equivalent width of H${\beta}_{\rm BC}$ and that of the Fe$_{\rm II}$ 
complex centered at $\lambda4570$ (see e.g. Marziani et al. 2003; Sulentic et 
al. 2004 and references therein). Unfortunately, due to the limitations of
the 2dF spectra we cannot accurately measure the Fe$_{\rm II}$ 
complex and test for the above effect.  However, our composite 
spectrum for radio-loud sources does appear remarkably similar to those of the 
Marziani et al. B1 population (Marziani, private communication), giving us 
some confidence that the majority of our radio-loud quasars falls in this 
category and therefore constitutes a roughly homogeneous sample also in the E1 
space.

\section{Fitting Line Widths}\label{sec:fitting-line-widths}

\begin{figure}
\centering\epsfig{figure=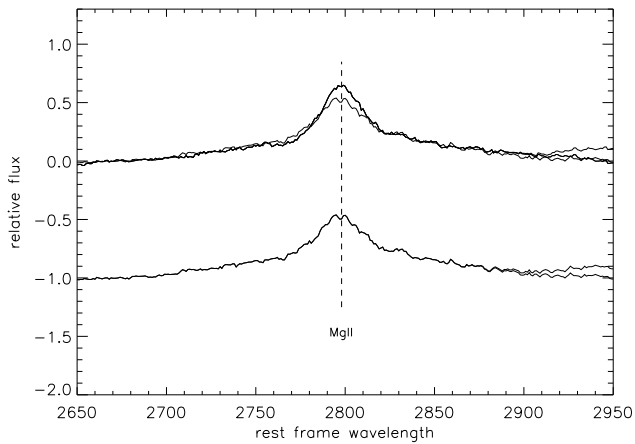,height=2.5in}
\caption[]{Blowup of the spectra shown in
  figure~\ref{fig:composite} around the \mgii line.  The thick upper curves
  are the composite spectra for RQ sources with and without the
  contaminating lines subtracted, in this case \feii.  The thick curves
  on the bottom are the same spectra for the RL quasars.  
  The thin spectrum at the top is the RL spectrum reproduced and shifted up for
  better comparison with the RQ spectrum.}
\label{fig:composite_MgII}
\end{figure}

\begin{figure}
\centering\epsfig{figure=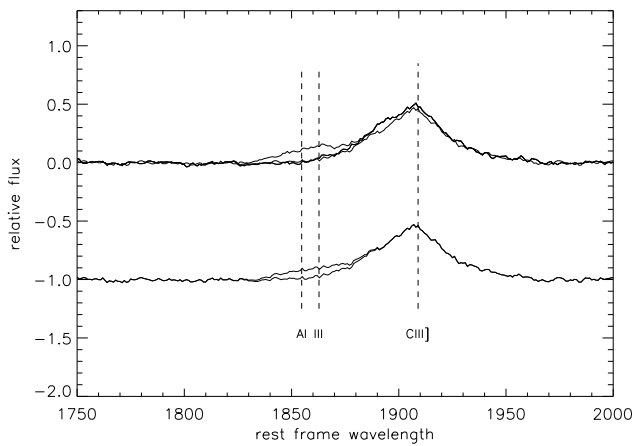,height=2.5in}
\caption[]{Blowup of the spectra shown in
  figure~\ref{fig:composite} around the \ciii line.  The identities of
  the spectra are the same as in figure~\ref{fig:composite_MgII}.
  There is also a \siiii line at 1892\AA~ that was not subtracted.
 Note that the removal 
  of the \aliii lines does
  not significantly change the FWHM of the \ciii + \siiii lines, which are 
 found to 
  be very similar in the two RL and RQ cases (see Table~1).}
\label{fig:composite_CIII}
\end{figure}

\begin{figure}
\centering\epsfig{figure=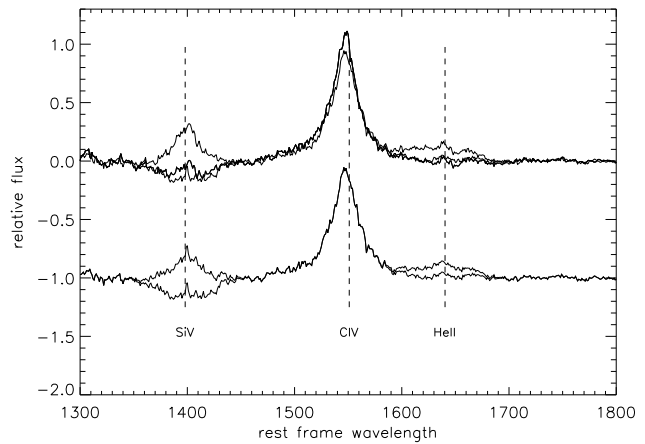,height=2.5in}
\caption[]{Blowup of the spectra shown in
  figure~\ref{fig:composite} around the \civ line.  The identities of
  the spectra are the same as in figure~\ref{fig:composite_MgII}.  The
  blueshift of these lines relative to \mgii is
  evident when one compares the positions of these emission lines in the
  restframe (as marked by the dashed lines).  There is no significant
  difference in the width of the RL and RQ \civ lines (see Table~1).}
\label{fig:composite_CIV}
\end{figure}

The widths of the \mgii, \ciii and \civ lines are measured from the
composite spectra in figure~\ref{fig:composite} and the luminosity-binned 
spectra in figures~\ref{composite_RQ} and \ref{composite_RL}.
The widths of the \hb and \lya lines are not used here because we find the 
noise in
these two regions of the composite spectra to be too large to make useful
measurements.  The regions around the \mgii, \ciii and \civ lines are shown 
in more detail in figures~\ref{fig:composite_MgII}, \ref{fig:composite_CIII}
and \ref{fig:composite_CIV}.

Contaminating lines are
removed before the widths of the three major lines are measured.  
In each case, the main line and the contaminating lines are fit to a
model and then the model for the contaminating lines alone is
subtracted from the spectrum.  Each contaminating line is modeled as
a Gaussian and the main lines -- \mgii, \ciii and \civ -- are modeled as
two Gaussians, one with a broad and one with a narrow component.  
The relative centers of the lines are
fixed, but the redshift is allowed to vary during the fit.  

Note that there is a \feii feature longword of the \mgii line (see 
figures~\ref{fig:composite} and \ref{fig:composite_MgII}).  
This feature is subtracted, but
does not have any significant effect on the measured line width
because \feii is well out in the wing of the \mgii line.
An \alii doublet is subtracted from
the \ciii line (see figure \ref{fig:composite_CIII}).  
\siv and \heii are removed from the \civ line 
(figure \ref{fig:composite_CIV}).  None of these lines have a 
significant effect on the measured FWHM line widths.

We tested several methods for measuring the line widths and found them
all to be consistent.  One was to use the FWHM of the model that was
fit to the data in order to remove the contaminating lines.  This
method assumes that the line shape is similar to the double Gaussian
used to model it. 
Another method was to median smooth the spectrum over 5 pixels and
then find the FWHM.  The widths reported in this paper were calculated
by first finding the maximum pixel value within the line.  To reduce
the dependence on noise in a single pixel, we take the median of the
maximum pixel value and its four nearest neighbors.  Using this as the
peak value of the line we find the closest pixel on each side of the
line with a value below half the peak value.
It was found that this last method has the least variance and makes a
minimum of assumptions about the shape of the line while agreeing very
well with the other two methods.
We also calculated an inter-percentile velocity (IPV) width by
integrating the spectra around the peak of the
line and finding the region symmetrically surrounding the peak that
contains 76\% of the total (corresponding to the area within the FHWM of 
the Gaussian).  
We found that in comparison to the FWHM the IPV was
more sensitive to the wings of the line and more strongly dependent on the 
subtraction 
of the contaminating lines. We therefore discarded it in favor of the FWHM.
Table~\ref{tbl:line_widths} shows the FWHM measures of the 
line widths for the radio-loud and radio-quite populations. The values
derived for our sample of radio-quiet quasars are in very good  
agreement with those obtained by Corbett et al. (2003).  Note that the
entire broad line width is used; no attempt is made to separate narrow
and broad components.

Errors in the line widths are derived by the bootstrap method. Spectra are 
drawn at
random from the samples of continuum-normalized spectra with
replacement -- 352 for the radio-quite population and 291 for the radio-loud 
one. From these a composite spectrum is constructed in the same way as
discussed in \S~\ref{sec:constr-comp-spectra} and the line widths are
measured as discussed above.  This process is repeated 1000 times.
The errors given in table~\ref{tbl:line_widths} are
the 68 percentile (corresponding to 1-sigma) confidence regions
derived from the resulting distributions.

\begin{table*}
\begin{center}
\caption{ Widths of the  \mgii, \ciii  and \civ broad lines.  
 These are all consistent for radio-loud and radio-quiet sources except 
 in the case of \mgii.
\label{tbl:line_widths}}
\begin{tabular}{lccc}
  & \mgii & \ciii & \civ \\
\hline
\\
Radio-loud (FWHM) &  
$4710^{+210}_{-220}\kms$ & 
$5810^{+310}_{-320}\kms$ & 
$5410^{+520}_{-440}\kms$ 
\\
Radio-quiet (FWHM) &  
$3960^{+250}_{-290}\kms$ &  
$5970^{+270}_{-200}\kms$ & 
$5020^{+390}_{-390}\kms$ 
\end{tabular}

\end{center}
\end{table*}

\section{Results}\label{sec:results}

Table~\ref{tbl:line_widths} shows that of the three lines analyzed in our work 
only the \mgii line has a width that is measurably correlated with
radio activity while the \ciii and \civ line widths are consistent
with being the same in the RL and RQ samples.
This is perhaps not too surprising given that \mgii is a low-ionization
line that is believed to originate in virialized gas clouds, which in turn
implies that, like for H$\beta$, its line width can be considered as a good 
tracer of the quasar's black hole mass (McLure \& Jarvis 2002). 
The same cannot be said of \civ since there is strong evidence for a 
systematic blueshift of this line indicating
that it is more likely to be associated with some form of outflow
rather than produced by  virialized clouds (e.g. Marziani et al. 1996;
Bachev et al. 2004).

The blueshift of the \civ lines can be clearly seen in
figure~\ref{fig:composite_CIV}.  The relative blueshift between the
peaks of the \mgii and \civ lines was measured in the composite
spectra and the bootstrap realizations.  The relative shifts were
found to be $-400^{+300}_{-400}\kms$ for the RL sample and
$-600^{+200}_{-200}\kms$  for the RQ sample.  A larger \civ
blueshift in the RQ quasars is in agreement with previous findings
(e.g. Marziani et al. 1996, Richards et al. 2002 and Bachev et
al. 2004).

In contrast to \civ, the \ciii lines in the RL and RQ composite spectra are 
remarkably
similar although in this case the \siiii line -- that we did not attempt
to subtract (figure~\ref{fig:composite_CIII}) -- makes the interpretation 
difficult.  It does seem as though the only
differences between the RL and RQ line profiles in the case of \ciii 
are attributable to differences in the \siiii lines.


We use the McLure \& Jarvis (2002) estimator to derive the average black 
hole mass for the different samples of radio-loud and radio-quiet quasars.
Under the assumption that the motion of the \mgii-emitting material is 
virialized, one can estimate the central black hole mass via the expression:
\begin{eqnarray}
{\rm M_{BH}}=G^{-1}{\rm R_{[\mgii]}}{\rm V_{[\mgii]}}^2,
\end{eqnarray}
where $\rm R_{[\mgii]}$ is the radius of the broad-line region and 
$\rm V_{[\mgii]}$ is the Keplerian velocity of its gas, which can be written 
as $f\cdot {\rm FWHM_{[\mgii]}}$. The geometric factor $f$  
relates the FWHM to the intrinsic Kaplerian velocity of the emitting gas in 
the \mgii region. Assuming random isotropic orbits, 
$f=\sqrt{3}/2$ (see McLure \& Jarvis 2002 and McLure \& Dunlop 2004 for 
further considerations on this issue).
To relate the radius of the \mgii region to the luminosity 
at 3000\AA~ we use the steeper relation derived by McLure \& Dunlop
(2004) for high-luminosity (SDSS) quasars, such as those considered in this 
analysis,
 instead of the fit given in McLure \& Jarvis (2002). We therefore adopt:
\begin{eqnarray}\label{eq:size_matters}
\rm R_{[\mgii]}=(18.5\pm 6.6)\;[\lambda L_{3000}/10^{37}{\rm W}]
^{(0.62\pm 0.14)}.
\end{eqnarray}

With the above relations, the black-hole mass estimator can then be written as:
\begin{eqnarray}
\frac{\rm M_{BH}}{M_\odot}=f\cdot 3.2 \;\left(\frac{\lambda{\rm L_{3000}}}
{10^{37}{\rm W}}\right)^{0.62}\left(\frac{\rm{FWHM_{\rm \mgii}}}
{\rm{kms^{-1}}}\right)^2,
\end{eqnarray} 
with ${\rm L_{bol}}=5.9 \cdot \lambda {\rm L_{3000}}$, and 
\begin{eqnarray}\label{eq:mag_lum}
{\rm M_B}=-2.66(\pm 0.05)\;{\rm log[L_{bol}/W]}+79.36(\pm 1.98),
\end{eqnarray}
where ${\rm L_{bol}}$ is the bolometric luminosity and ${\rm M_B}$ is the 
absolute B magnitude (see McLure \& Dunlop 2004 for further details).

Our mass estimates are derived from equations (\ref{eq:size_matters})
through (\ref{eq:mag_lum}) using 
the  median values of the B magnitudes for the radio-loud 
and radio-quiet samples in equation~(\ref{eq:mag_lum}).  Note that these 
median values have not been calculated for the entire sample but  
only for those sources that contribute to the composite \mgii line, i.e. 237 
radio-loud and 220 radio-quiet quasars.  The average black hole masses 
for the two populations of radio-loud and radio-quiet quasars then read:
\begin{eqnarray}\label{eq:Mbh_all}
{\rm log \left(\frac{M_{BH}}{M_\odot}\right)_{RL}}=8.56\pm 0.04
\\
{\rm log \left(\frac{M_{BH}}{M_\odot}\right)_{RQ}}=8.30\pm 0.06,
\nonumber
\end{eqnarray}
which shows with a confidence level $\simgt 3\sigma$ that black 
holes associated with radio-loud quasars are in general 
about twice as massive as those producing the radio-quiet case.

\begin{figure}
\centering\epsfig{figure=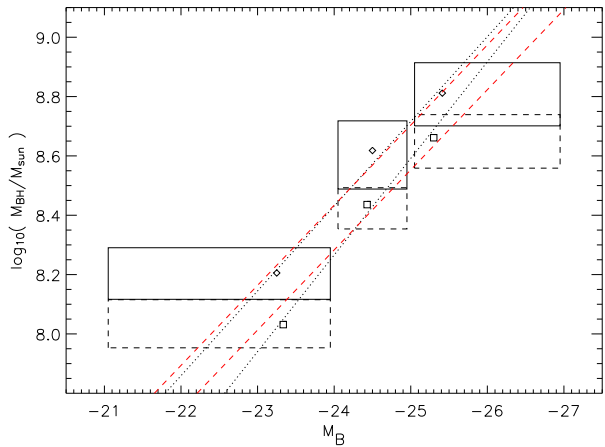,height=2.5in}
\caption{The black hole mass derived from \mgii line widths in
  different B-band absolute magnitude bins.  The boxes show the range of the 
  $\rm M_B$ bins
  and the one sigma error bars in $\log_{10}\rm M_{\rm BH}$.  The solid
  boxes are for the radio-loud sample and the dashed boxes are for
  radio-quite sample.  The median values are shown by squares and
  diamonds.  The dotted lines are the best fits to the data, while the 
  dashed lines show the fits when only normalization is allowed to 
  vary (see section~\ref{sec:results}).}
\label{MBHvsM_fwhm}
\end{figure}

\begin{figure}
\centering\epsfig{figure=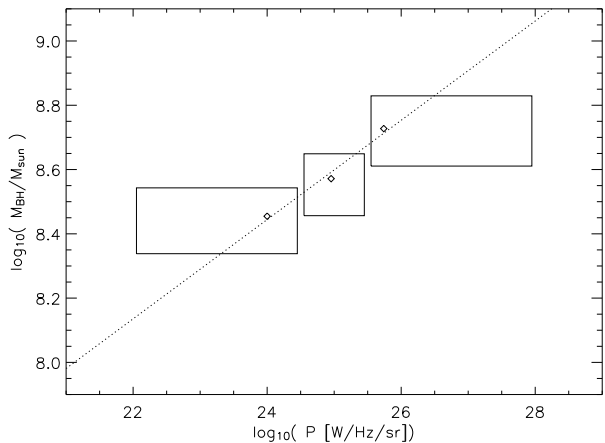,height=2.5in}
\caption{The black hole mass as derived from \mgii line widths in
  different bins of radio power. As in figure \ref{MBHvsM_fwhm}, the 
  boxes show the range of the $\rm {\log_{10}\;P\;[W/Hz/sr]}$ bins
  and the one sigma error bars in $\log_{10}\rm M_{\rm BH}$. The dotted line 
  is the best fit to the data (see section~\ref{sec:results}).}
\label{MBHvsP}
\end{figure}

To further investigate this difference we examine the line widths and
black hole masses for the QSO samples binned in absolute B-band magnitudes
(figures~\ref{composite_RQ} and \ref{composite_RL}).
Figure~\ref{MBHvsM_fwhm} shows the black holes masses in
these bins.  Linear fits to the data give
\begin{eqnarray}
{\rm log \left(\frac{M_{BH}}{M_\odot}\right)_{RL}}= 8.57(\pm0.06) -
0.27(\pm 0.06) ({\rm M_B} + 24.5) \\
{\rm log \left(\frac{M_{BH}}{M_\odot}\right)_{RQ}}= 8.43(\pm0.05) -
0.32(\pm 0.06) ({\rm M_B} + 24.5) 
\end{eqnarray}
The RL and RQ normalizations are inconsistent with each other at the two 
sigma level and agree
with the values derived from using the whole luminosity range 
(\ref{eq:Mbh_all}). On the other hand, the slopes are consistent with being 
equal and indicate that the two populations follow a very similar 
relationship where the black hole mass is almost linearly dependent on 
luminosity, $\rm {M_{BH}}\propto {\rm {L_B}^{0.7}}$ (see 
equation~\ref{eq:mag_lum}).

Corbett, et al.(2003), using a much larger sample from 2dF QSO survey,
found a relation between the \mgii line width and B-band luminosity that
is given by $\log(v)=1.43(\pm0.4) + 0.048(\pm 0.009)\log(L_b)$ (no
distinction was made between RL and RQ QSOs).  This
translates into a slope of $0.27$ in the relations above, value which is  
within 1-sigma of both of our RL and the RQ slopes.  We can therefore justifiably
assume that the slope is the same for RL and RQ QSOs and fit for just the
normalization, or typical black hole mass.  These fits read
\begin{eqnarray}
{\rm log \left(\frac{M_{BH}}{M_\odot}\right)_{RL}}= 8.57(\pm0.06) -
0.27 (\rm M_B + 24.5) \\
{\rm log \left(\frac{M_{BH}}{M_\odot}\right)_{RQ}}= 8.42(\pm0.04) -
0.27 (\rm M_B + 24.5), 
\end{eqnarray}
and indicate -- together with equations (6) and (7) -- that at all 
luminosities RL QSOs show a systematic trend to be more massive than RQ ones.
More specifically one has that for a given luminosity, a typical radio-loud 
QSO has a black hole mass
that is about $1.4$ times as massive as a radio-quite QSO of the same 
B-band luminosity. Note that this result could also be interpreted as evidence 
for the hypothesis that, at each given luminosity, there is a critical black 
hole mass that is required to produce relevant radio activity. 

The above results are in good agreement with the findings of McLure \& Jarvis 
(2004) who analysed two samples of radio-loud and radio-quiet quasars drawn from the 
SDSS Quasar Catalog II (Schneider et al. 2003) so to be indistinguishable in terms 
of redshift and optical luminosity distributions. The above authors in fact conclude 
that the average black hole mass associated to radio-loud quasars is about twice that 
of radio-quiet quasars. They also report that such black hole mass difference is constant 
over the full luminosity range covered by their samples. 
Jarvis \& McLure (2004) find black hole masses that are slightly greater than those 
derived in this work, but the difference is most likely to be attributed to the different 
selection criteria of the 2dF and SDSS surveys, the latter one preferentially selecting 
brighter quasars than 2QZ.

Another interesting result that can be obtained from our samples is the 
relationship between black hole mass and radio luminosity. The trend is 
illustrated in 
figure \ref{MBHvsP} for the population of radio-loud quasars. A linear fit to 
the data leads to
\begin{eqnarray}
{\rm log \left(\frac{M_{BH}}{M_\odot}\right)_{RL}}= 8.60(\pm0.06) +
0.15(\pm 0.08) ({\rm \log_{10}P} -25)
\end{eqnarray} 
(where the radio power is measured at 1.4 GHz and given in [W/Hz/sr] 
units), which shows that radio luminosity and black hole mass are very loosely 
correlated since the dependence -- if any -- between these two quantities 
is much weaker than in the black hole/optical luminosity case.\\
The above results can be interpreted as evidence for the fact that while 
optical/nuclear emission is directly related to the mass of the black 
hole that produces the quasar signal, radio luminosity is not. 
Black hole mass in RL quasars merely plays a marginal ``threshold'' role 
whereby, for a given optical luminosity, only the more massive quasars will 
develop significant radio activity. However, once the activity is 
triggered, there is no significant evidence for a
connection between black hole mass and the level of radio output 
(see also Magliocchetti et al. 2004).

\section{Conclusions}\label{sec:interpr--disc}

We have made use of a homogeneous sample of $\sim 300$,
$0.3\simlt {\rm z}\simlt 3$, radio-loud quasars drawn from the FIRST and 2dF
QSO surveys to estimate possible
dependences of radio activity on black-hole mass. By analyzing composite
spectra for the populations of radio-quiet and radio-loud QSOs -- chosen to
have the same redshift and luminosity distribution -- we find
with high statistical significance
that radio-loud quasars are on average associated to black holes of masses
$\sim 10^{8.56\pm 0.04}{\rm M_\odot}$, about twice as large as those obtained 
for radio-quiet quasars ($\sim 10^{8.30\pm 0.06}{\rm M_\odot}$).

The above result is also verified when one splits the two samples in 
luminosity bins. In fact, we observe a clear dependence of black hole mass 
on optical luminosity of the form ${\rm log \left(\frac{M_{BH}}
{M_\odot}\right)_{RL}}= 8.57(\pm0.06) -0.27(\pm 0.06) ({\rm M_B} + 24.5)$ and
${\rm log \left(\frac{M_{BH}}{M_\odot}\right)_{RQ}}= 8.43(\pm0.05) -
0.32(\pm 0.06) ({\rm M_B} + 24.5)$, respectively for the case of radio-loud
and radio-quiet quasars. These trends run
parallel to each other, implying that radio-loud quasars are associated to
black holes about $1.4$ times more massive than those producing the 
radio-quiet case {\it at all sampled luminosities}.
On the other hand, in the case of radio-loud quasars, we only find evidence
for a marginal (if any) dependence of the black hole mass on radio power 
(${\rm log \left(\frac{M_{BH}}{M_\odot}\right)_{RL}}= 8.60(\pm0.06) +
0.15(\pm 0.08) ({\rm \log_{10}P} -25)$).\\

\noindent
Our results allow a number of important conclusions to be drawn:\\
i) There exists a strong correlation between 
level of optical/nuclear emission and mass of the black hole producing the 
quasar. The same correlation holds for both radio-loud and radio-quiet
quasars, but with different zero-points.\\
ii) Radio-loud quasars are on average more massive than radio quiet ones at 
all (optical) luminosities.\\
iii) Black hole mass plays a partial, threshold, role in determining
the radio activity 
that gives rise to the population of radio-loud quasars. Our findings
are further support for
the existence -- at each given optical luminosity -- of a threshold black hole
mass associated with the onset of
significant radio activity such as that of radio-loud QSOs; however,
once the activity is triggered, there appears to be very little evidence for a
connection between black hole mass and level of radio output (see also 
Magliocchetti et al. 2004).\\


An important issue to stress is that, while the absolute values associated 
to the various quantities analyzed in this paper might be somehow affected by 
calibration effects (see e.g. section 5), all the results that compare 
the two populations of radio-loud and radio-quiet quasars are not, as any 
difference in the zero-point calibrations would cancel out during the 
comparison process. Even though the analyzed samples are too small to 
investigate for any presence of a different cosmological evolution of 
the two RQ and RL populations, the above statement, together with our choice 
of selecting two samples that were as``indistinguishable'' as possible 
(in number of sources, luminosity and redshift distributions), give us 
confidence in the reliability of the results. 

As a final remark we note that, since we have chosen two samples matched in 
optical luminosity, the fact that RL quasars harbor on average more 
massive black holes implies that these sources accrete at a lower fraction 
of their Eddington limit than their RQ counterparts. Whether this  
result can be seen as evidence that radio-loud and radio-quiet quasars 
are two distinct populations or whether this can simply be interpreted by 
saying that radio-loud quasars mark a different 
(possibly later) stage of evolution of radio-quiet quasars is still unclear 
and we plan to tackle this issue in a forthcoming paper.

\section*{Acknowledgments}
\footnotesize
The authors wish to thank Paola Marziani for extremely interesting discussions 
and Gianfranco De Zotti for a careful reading of the manuscript.
Financial support for RBM was provided by NASA through Hubble
Fellowship grant HF-01154.01-A awarded by the Space Telescope Science
Institute, which is operated by the Association of Universities for Research 
in Astronomy, Inc., for NASA, under contract NAS 5-26555. We also thank the anonymous 
referee to help improving the robustness of the results of this paper.


\appendix{}

\section{interpretation of composite spectra}

If the luminosity distribution of the radio-loud and radio-quite QSOs
are the same then the difference in their composite spectra must
reflect a difference in the spectra of radio-loud and radio-quite QSOs
{\it of the same luminosity}.  This is important because a difference
in line width can then be directly interpreted as a difference in
black hole mass.  If this were not the case, we might be comparing
radio-loud QSOs at one luminosity to radio-quite QSOs at another
luminosity and then the derived difference in black hole mass might
just be the result of this difference in luminosity.  In that case we
would not be able to conclude that radio-loud and radio-quite of the
same visible luminosity tend to have black holes of different masses.

Let the QSO spectra be binned into arbitrarily narrow bins in visible
luminosity $L$.  Let us say there are a total of $N$ QSOs in the
sample and there
are $n(L)$ QSOs in the luminosity bin centered on $L$.  The composite
spectrum $f_c(\lambda)$ is the average of these spectra at fixed
 wavelength.  The average difference between the composite
spectra of the radio-loud and radio-quite is
\begin{eqnarray}
\left\langle f_c^{\rm RL}(\lambda)-f_c^{\rm RQ}(\lambda) \right\rangle
\;\;\;\;\;\;\;\;\;\;\;\;\;\;\;\;\;\;\;\;\;\;\;\;\;\;\;\;\;\;\;\;\;\;\;\;
\;\;\;\;\;\;\;\;\;\;\;\;\;\;\;\;\;\;\;\;
\nonumber\\
 =\frac{1}{N_{\rm RL}} \sum_L \left\langle \sum_i  f^{\rm RL}_i(\lambda,L) 
\right\rangle - 
\frac{1}{N_{\rm RQ}} \sum_L \left\langle \sum_j  f^{\rm RQ}_j(\lambda,L)
\right\rangle \\
 = \sum_L \left\{ \frac{1}{N_{\rm RL}} \left\langle \sum_i f^{\rm
  RL}_i(\lambda,L) \right\rangle
      - \frac{1}{N_{\rm RQ}} \left\langle \sum_i f^{\rm RQ}_i(\lambda,L)
      \right\rangle \right\} \\
= \sum_L \left\{ \frac{n_{\rm RL}(L)}{N_{\rm RL}} \left\langle f^{\rm
  RL}(\lambda,L) \right\rangle
      - \frac{n_{\rm RQ}(L)}{N_{\rm RQ}} \left\langle f^{\rm RQ}(\lambda,L) 
\right\rangle \right\}\;\;\;\;\;\;\\
 = \sum_L \Phi(L) \left\{  \left\langle f^{\rm
  RL}(\lambda,L) \right\rangle
      - \left\langle f^{\rm RQ}(\lambda,L) \right\rangle \right\}
\;\;\;\;\;\;\;\;\;\;\;\;\;\;\;\;\;\;\;\;\;\;\;\;\;\;
\end{eqnarray}
where $\Phi(L)$ is the luminosity distribution.  In the last step the
requirement that $\Phi(L)$ is the same for radio-loud and
radio-quite was used.  On average, there will be no difference in the
composite spectra unless there is a systematic difference in the
spectra of QSOs of the same luminosity.

In the this paper we use the median instead of the mean to calculate
the composite spectra.  Using the mean gives consistent results, but
the noise is greater.  We feel that the noise is large compared to any
possible skew in the distribution of pixel values amongst QSO
spectra so the median and mean will give consistent results, but that
the median is less sensitive to stray outliers.

\end{document}